\documentclass[pdflatex,sn-mathphys-num]{sn-jnl}

\usepackage{graphicx}%
\usepackage{multirow}%
\usepackage{amsmath,amssymb,amsfonts}%
\usepackage{amsthm}%
\usepackage{mathrsfs}%
\usepackage[title]{appendix}%
\usepackage{xcolor}%
\usepackage{textcomp}%
\usepackage{manyfoot}%
\usepackage{booktabs}%
\usepackage{algorithm}%
\usepackage{algorithmicx}%
\usepackage{algpseudocode}%
\usepackage{listings}%
\usepackage{graphicx}
\usepackage{slashed}
\usepackage{comment}
\usepackage{braket}

\theoremstyle{thmstyleone}%
\theoremstyle{thmstyletwo}%

\theoremstyle{thmstylethree}%

\begin{document}

\title{Geometric Phases as Probes of Dark Sectors }
\author*[1,2]{\fnm{Antonio} \sur{Capolupo}}\email{acapolupo@unisa.it}
\equalcont{These authors contributed equally to this work.}
\author[1,2]{\fnm{Gabriele} \sur{Pisacane}}\email{gpisacane@unisa.it}
\equalcont{These authors contributed equally to this work.}
\author[2]{\fnm{Raoul} \sur{Serao}}\email{rserao@unisa.it}
\equalcont{These authors contributed equally to this work.}

\affil[1]{\orgdiv{Department of Physics}, \orgname{University of Salerno}, \orgaddress{\street{Via Giovanni Paolo II, 132}, \city{Fisciano}, \postcode{84084}, \state{}, \country{Italy}}}

\affil[2]{\orgdiv{INFN}, \orgname{gruppo collegato di Salerno}, \orgaddress{\street{Via Giovanni Paolo II, 132}, \city{Fisciano}, \postcode{84084}, \state{}, \country{Italy}}}


\abstract{
We review recent interferometric schemes designed to probe physics beyond the Standard Model through the detection of geometric phases. We discuss how interactions with hidden-sector degrees of freedom, such as axion-like particles and mirror-matter candidates, can induce potentially observable phase shifts in ordinary fermion systems \cite{Capolupo1,CapolupoAx2021,bib1}.
}




\maketitle

\section{Introduction}\label{sec1}

The Standard Model (SM) of particle physics provides an extremely successful
description of fundamental interactions, but several theoretical and experimental
indications show that it cannot be regarded as complete. Neutrino oscillations
\cite{Neut1,Neut3,Neut4,Neut5,Neut6,Neut7,Neut8,Neut9,Neut10} provide direct
evidence for physics beyond the minimal SM, since they require non-vanishing
neutrino masses and flavor mixing. At the same time, precision anomalies such as
the muon anomalous magnetic moment \cite{Muon1,Muon2}, together with the strong
astrophysical and cosmological evidence for dark matter and dark energy
\cite{Rubin1,Rubin2,Trimble1987,Corbelli2000,Clesse2018,Capolupo2021,Salucci2021,
Capolupo2020,Capolupo2025,Quaranta1,Pisacane1}, motivate the search for new
particles, new interactions, and possible departures from exact fundamental
symmetries.

Among the most widely studied extensions of the SM are scenarios involving
weakly coupled particles or hidden sectors. Axions and axion-like particles,
originally introduced in connection with the strong CP problem and later developed
as compelling dark-matter candidates
\cite{Peccei1,Peccei2,Weinberg,Wilczek,Raffelt,CapolupoAx2020,Marsh}, represent a
paradigmatic example of light bosonic degrees of freedom that may couple feebly to
ordinary matter. Another possibility is provided by mirror-matter models, in which
ordinary particles may mix with hidden-sector partners
\cite{YangLee1956,Kobzarev1966,Blinnikov1982,Berezhiani2004,Hodges1993,Foot2014,
Hostert2023,Hao2022,Berezhiani2006,Berezhiani2019,Czarnecki}.

A common difficulty in all these searches is that the expected effects are
typically very small. For this reason, high-precision quantum interferometric
techniques provide a natural tool to probe weak interactions, tiny phase shifts,
and subtle modifications of quantum dynamics. Neutron interferometry
\cite{RauchBook,Werner1979,Wagh1990,Allman1997,Filipp2005} and atomic
interferometry \cite{brax} have long been used to investigate particle
interactions, quantum coherence, and phase effects. In such setups, even weak
interaction-induced corrections can become observable through their contribution
to interference patterns.

In this context, geometric phases \cite{Berry,Berry1,Mukunda} offer a particularly useful class of observables.
 
In this work, we review recent results showing how interactions with hidden-sector degrees of freedom can modify the phase evolution of ordinary quantum systems and generate observable interferometric signatures.  We show that fermion–fermion interactions mediated by axions and axion-like particles can be probed through interferometric measurements of neutron beams \cite{CapolupoAx2021}. We further demonstrate that neutron interferometry can probe mirror-neutron scenarios, since neutron–mirror-neutron mixing would generate a geometric phase in the ordinary neutron state that could be detected interferometrically \cite{bib1}.

\section{Geometric Phases and Axion-like particles}
\label{sec:dark_sector}
We employ the kinematic formulation of the non-cyclic and non-adiabatic geometric
phase introduced in Ref.~\cite{Mukunda}, which is particularly well suited to the
interferometric configurations considered below. For a normalized quantum state
$|\psi(t)\rangle$, the geometric phase is  defined as the difference between the total phase and the dynamical phase,
$
\Phi^{g}(t)=\Phi^{\rm tot}(t)-\Phi^{\rm dyn}(t),
$
with
$
\Phi^{\rm tot}(t)=\arg\langle \psi(0)|\psi(t)\rangle,
$ and $
\Phi^{\rm dyn}(t)=
\Im\int_{0}^{t}\langle \psi(\tau)|\dot{\psi}(\tau)\rangle\,d\tau .
$
 
Besides their coupling to photons, which is the basis of many
experimental searches, axions and axion-like particles (ALPs) can also couple to fermions and mediate effective
spin-dependent interactions between ordinary matter particles. In the neutron
sector, this interaction can modify the phase accumulated during propagation and
therefore produce an additional contribution to the interference pattern of a
neutron interferometer.
We consider the  interaction between the axion field and
fermions \cite{CapolupoAx2021},
$\mathcal{L}_{\rm INT}= -\sum_{j=1,2} i g_{aj}\,\varphi\,\bar{\psi}_{j}\gamma_{5}\psi_{j},$
where $\varphi$ denotes the axion field, $\psi_j$ are fermion fields, and $g_{aj}$
are the effective axion--fermion couplings. For neutron--neutron interactions one
has $g_{aj}=g_{aNN}\equiv g_p$, with $g_p$ denoting the effective axion--neutron
coupling constant.

In the non-relativistic regime, axion exchange induces a spin-dependent
interaction between neutrons \cite{CapolupoAx2021}. Assuming inter-neutron distances
$r>10^{-12}\,\mathrm{m}$, so that short-range nuclear forces can be neglected, the
relevant two-body Hamiltonian contains both the ordinary magnetic dipole--dipole
interaction and the ALP-mediated contribution:
\begin{equation}
H_{ij}
=
-\frac{\mathcal{A}}{r_{ij}^{3}}
\left[
\left(
3-\mathcal{B}e^{-m r_{ij}}K^{(a)}(r_{ij})
\right)
\sigma_{i}^{r_{ij}}\sigma_{j}^{r_{ij}}
-
\left(
1-\mathcal{B}e^{-m r_{ij}}K^{(b)}(r_{ij})
\right)
\boldsymbol{\sigma}_{i}\cdot\boldsymbol{\sigma}_{j}
\right].
\end{equation}
Here
$
\mathcal{A}=\frac{g^{2}\alpha}{16M^{2}},
\mathcal{B}=\frac{g_{aNN}^{2}}{\pi\alpha g^{2}}
=
\frac{g_{p}^{2}}{\pi\alpha g^{2}}, $
where $M$ is the neutron mass, $g$ is the neutron $g$-factor, $\alpha$ is the
fine-structure constant, and $m$ is the ALP mass. The functions appearing in the
Yukawa-suppressed part of the interaction are
$K^{(a)}(r)=m^{2}r^{2}+3mr+3,
K^{(b)}(r)=mr+1.$
The parameter $\mathcal{B}$ measures the relative strength of the axion-induced
interaction. Therefore, when $g_{aNN}=0$, the Hamiltonian reduces to the ordinary
magnetic dipole--dipole case.

\begin{figure}[htpb]
	\centering
	\includegraphics[width=0.6\linewidth]{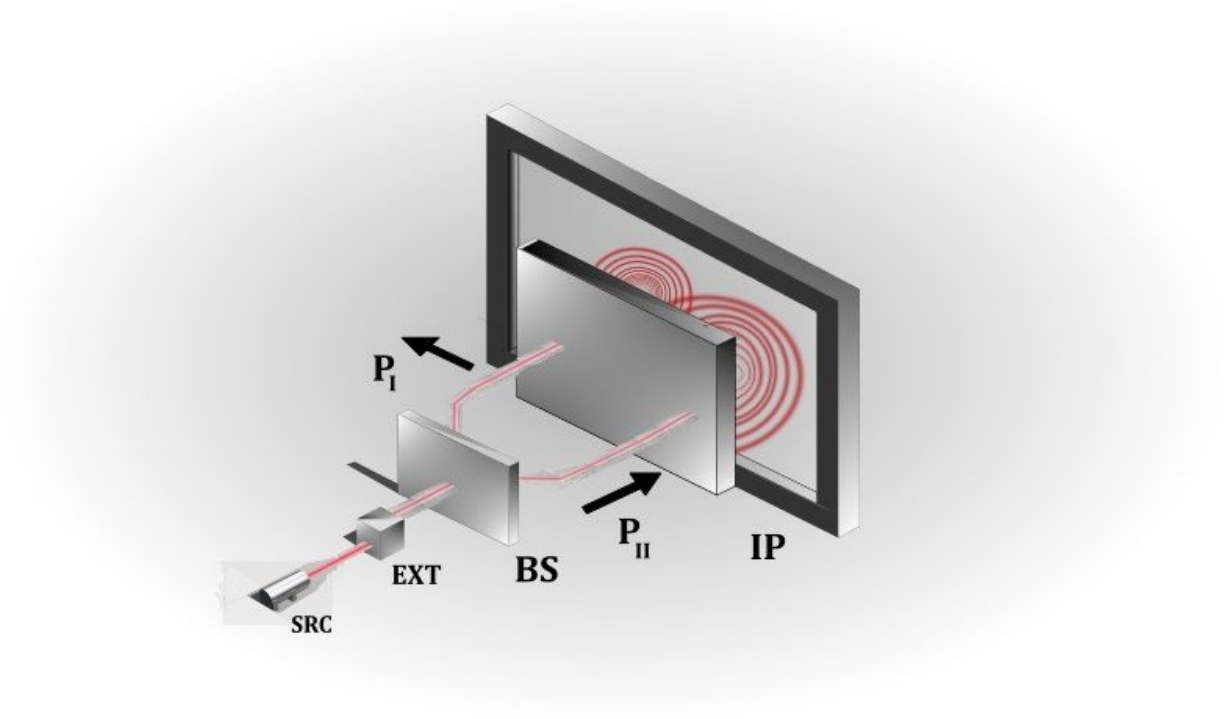}
	\caption{Schematic representation of the neutron interferometric setup.}
	\label{fig:interferometer}
\end{figure}

To connect this microscopic interaction with an interferometric signal, the
many-neutron problem is reduced to an effective one-particle description through a
mean-field approximation. The interaction of a neutron with the other neutrons in
the beam is encoded into an effective magnetic field. In the idealized setup shown
in Fig.~\ref{fig:interferometer}, a coherent cold-neutron beam is split into two
sub-beams, denoted by $I$ and $II$. The two beams are taken to have the same mean
inter-neutron distance,
$d_I=d_{II}=d,$
but they propagate through external magnetic fields of equal magnitude and
different orientation. One polarization is chosen parallel to the propagation
direction, while the other is chosen orthogonal to it.
Solving the effective single-neutron Schr\"odinger equation gives a relative phase
between the two sub-beams of the form \cite{CapolupoAx2021}
$\Delta\phi(t)=\left[G_m(d)+G_a(d)\right]t.$
The first term is the standard magnetic dipole--dipole contribution,
$G_m(d)=\frac{6\mathcal{A}}{d^{3}}\zeta(3),$
whereas the ALP-induced contribution is $
G_a(d)=-\frac{6\mathcal{A}\mathcal{B}}{d^{3}}
\mathrm{Li}_{3}\!\left(e^{-md}\right)
-\frac{6\mathcal{A}\mathcal{B}m}{d^{2}}
\mathrm{Li}_{2}\!\left(e^{-md}\right)
+
\frac{2\mathcal{A}\mathcal{B}m^{2}}{d}
\log\!\left(1-e^{-md}\right).$
Here $\zeta(s)$ is the Riemann zeta function and $\mathrm{Li}_{s}(z)=\sum_{n=1}^\infty \frac{z^n}{n^s}$ denotes the
polylogarithm. This decomposition makes explicit the physical content of the
calculation: the ordinary magnetic interaction and the axion-mediated interaction
contribute additively to the relative phase.

\begin{figure}[htpb]
	\centering
	\includegraphics[width=0.37\linewidth]{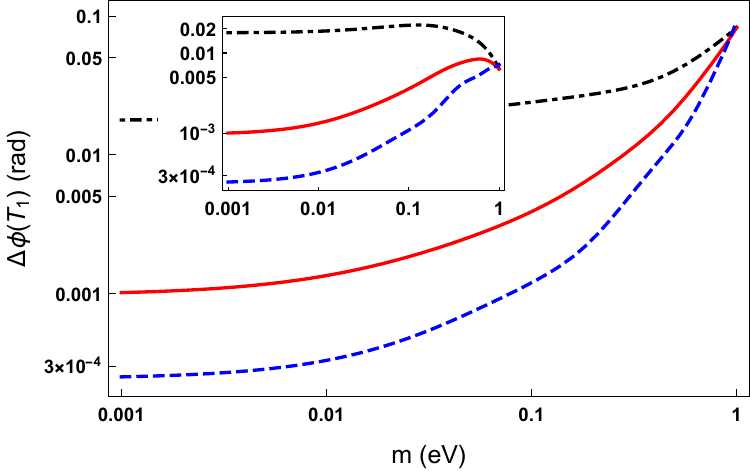}
	\caption{Logarithmic scale plot of  $|\Delta\phi(T_{1})|\bmod 2\pi$ as a function of the
	ALP mass in the range $[10^{-3},1]\,\mathrm{eV}$, for several values of the
	effective axion--neutron coupling $g_p\equiv g_{aNN}$ and for an inter-neutron
	distance $d=10^{-8}\,\mathrm{m}$ \cite{CapolupoAx2021}. The inset shows the corresponding result for
	$d=10^{-6}\,\mathrm{m}$. For small $md$, the phase is mainly controlled by the
	coupling, while the mass dependence becomes visible when Yukawa suppression is
	relevant.}
	\label{fig:Moste}
\end{figure}

\begin{figure}[htpb]
	\centering
	\includegraphics[width=0.48\linewidth]{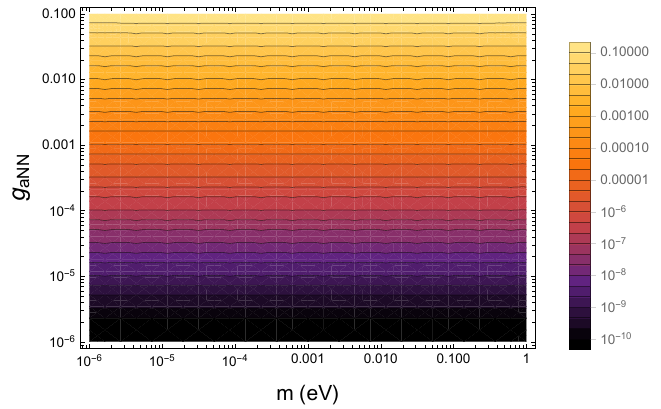}
	\hfill
	\includegraphics[width=0.48\linewidth]{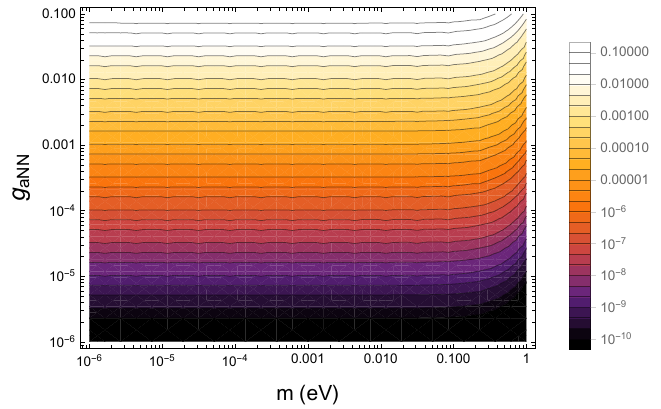}
	\caption{Contour plots of $|\Delta\phi(T_{1})|\bmod 2\pi$ in the
	mass--coupling plane
	$(m,g_p)\in[10^{-6},1]\,\mathrm{eV}\times[10^{-6},10^{-1}]$, for
	inter-neutron distances $d=10^{-9}\,\mathrm{m}$ in the left panel and
	$d=10^{-6}\,\mathrm{m}$ in the right panel. The bending of the contours at large
	$md$ reflects the Yukawa suppression of the ALP-mediated interaction.}
	\label{fig:Contour}
\end{figure}

The key step is to isolate the ALP contribution. This is done by choosing the
propagation time to coincide with the recurrence time
$
T_k=\frac{k\pi d^{3}}{3\mathcal{A}\zeta(3)},
\qquad k\in\mathbb{Z}.
$
At this time, the ordinary magnetic phase satisfies
$
G_m(d)T_k=2\pi k,
$
and is therefore invisible in the interference pattern modulo $2\pi$. The remaining
phase is then entirely due to the ALP-mediated neutron--neutron interaction:
\begin{equation}
\Delta\phi(T_k)
=
\left\{
\frac{k\pi\mathcal{B}}{3\zeta(3)}
\left[
2m^{2}d^{2}\log\!\left(1-e^{-md}\right)
-6md\,\mathrm{Li}_{2}\!\left(e^{-md}\right)
-6\,\mathrm{Li}_{3}\!\left(e^{-md}\right)
\right]
\right\}_{\rm mod\,2\pi}.
\end{equation}
Notice that $\Delta \phi(T_k)=0$ when $\mathcal{B}=0$, and therefore when the axion--neutron interaction is
absent. Conversely, a non-zero residual phase represents a signature of the ALP-mediated interaction.
A useful limit is obtained for $md\ll1$, which includes very light ALPs and small
inter-neutron distances. In this regime, the Yukawa damping is negligible and one
finds
$
\Delta\phi(T_k)
\simeq
\left\{-2k\pi\mathcal{B}\right\}_{\rm mod\,2\pi}.
$
The phase then becomes essentially independent of the ALP mass and of the
inter-neutron distance, and depends mainly on the coupling through
$\mathcal{B}\propto g_{aNN}^{2}$. This explains the behavior shown in
Fig.~\ref{fig:Moste}: for sufficiently small $md$, the curves are mainly separated
by the value of the coupling, while the mass dependence becomes visible only when
the Yukawa factor $e^{-md}$ starts suppressing the interaction \cite{CapolupoAx2021}.
The same conclusion is visible in the contour plots of Fig.~\ref{fig:Contour}. For
small inter-neutron distances, the phase contours are almost horizontal over a wide
mass range, showing that the signal is mostly controlled by the axion--neutron
coupling. At larger values of $md$, the contours bend because the finite ALP mass
suppresses the effective interaction. In this sense, neutron interferometry provides a
direct phase-based probe of the fermion--fermion interaction mediated by ALPs,
complementary to searches relying on the axion--photon coupling.

\section{Mirror matter and neutron mixing}

The hypothesis of a hidden gauge sector was first proposed in Ref.\cite{YangLee1956} as a
possible mechanism for restoring parity symmetry. A particularly well-studied
realization is provided by mirror-matter models \cite{Berezhiani2004,Hodges1993,Foot2014,Hostert2023,Hao2022,Berezhiani2006}, in which the hidden
sector possesses a particle spectrum structurally identical to that of ordinary
matter. The two sectors necessarily communicate through gravity and may also be
coupled by weak portal interactions, whereas strong and electromagnetic
interactions act separately within each sector. For neutrons, this possibility is described by neutron--mirror-neutron
mixing, $n$--$n'$, which may provide a low-energy signature of mirror matter and
can be probed through neutron interferometry \cite{bib1}.

Let $\ket{n}$ and $\ket{n'}$ denote the ordinary-neutron and mirror-neutron
states. The mixing relations are:
\begin{equation}
\begin{pmatrix}
\ket{n} \\
\ket{n'}
\end{pmatrix}
=
\begin{pmatrix}
\cos\theta & \sin\theta \\
-\sin\theta & \cos\theta
\end{pmatrix}
\begin{pmatrix}
\ket{n_1} \\
\ket{n_2}
\end{pmatrix}
\end{equation}
where $\ket{n_1}$ and $\ket{n_2}$ are the free Hamiltonian eigenstate.
For a fixed spin polarization, the effective Hamiltonian in the
$\{\ket{n},\ket{n'}\}$ basis is
$
H=
\begin{pmatrix}
m_n+\Delta E & \epsilon_{nn'} \\
\epsilon_{nn'} & m_n+\delta m
\end{pmatrix},
$
where $m_n$ is the neutron mass, $\epsilon_{nn'}$ is the mixing amplitude, and
$
\delta m=m_{n'}-m_n
$
is the mass splitting between mirror and ordinary neutron. The term $\Delta E$
accounts for the energy shift induced by external fields. In the interferometric
setup considered below, the relevant contribution is the coupling of the neutron
magnetic dipole to an external magnetic field. For spin aligned with the magnetic
field, we write
$
\Delta E = |\mu_n B|,
$
where $\mu_n$ is the neutron magnetic moment.

Diagonalization of the Hamiltonian gives the mixing angle
$
\tan(2\theta)=\frac{2\epsilon_{nn'}}{\delta m-\Delta E},
$
and the two eigenvalues
$
m_{1,2}
=
\frac{1}{2}
\left[
2m_n+\Delta E+\delta m
\mp
\sqrt{(\Delta E-\delta m)^2+4\epsilon_{nn'}^{\,2}}
\right].
$
For a neutron with momentum $k$, the corresponding energies are
$
\omega_j=\sqrt{k^2+m_j^2}, j=1,2,
$
and we define
$
\Delta\omega=\omega_2-\omega_1.
$
After removing an irrelevant common phase, the time-evolved neutron state can be
written as
$
\ket{n(t)}
=
\cos\theta\,e^{i\frac{\Delta\omega}{2}t}\ket{n_1}
+
\sin\theta\,e^{-i\frac{\Delta\omega}{2}t}\ket{n_2}.
$
Then, the geometric phase is \cite{bib1}:
$
\Phi^{g}(t)
=
\arg\!\left[
\cos\!\left(\frac{\Delta\omega}{2}t\right)
+
i\sin\!\left(\frac{\Delta\omega}{2}t\right)\cos(2\theta)
\right]
-
\frac{\Delta\omega t}{2}\cos(2\theta),
$
it is zero in the absence of mixing $\epsilon_{nn'}\rightarrow0$ and
$\theta\rightarrow0$. Therefore, a
non-zero geometric phase provides a signature of neutron--mirror
neutron mixing.

\begin{figure}[htpb]
	\centering
	\includegraphics[width=0.5\linewidth]{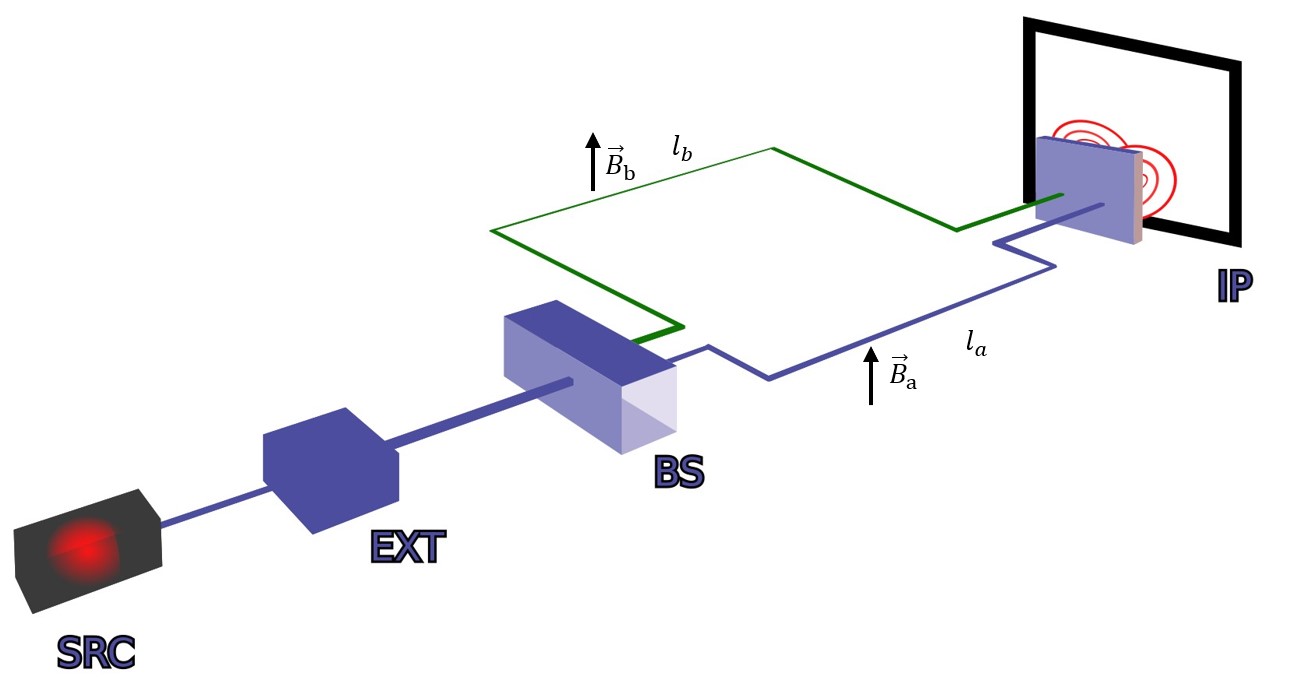}
	\caption{Schematic representation of the interferometric setup proposed to probe
	neutron--mirror-neutron mixing. The neutron beam is produced by the source
	\textbf{SRC}, polarized by the external device \textbf{EXT}, and then split by
	the beam splitter \textbf{BS}. The two sub-beams propagate along arms with
	different lengths and magnetic fields before being recombined at the
	interference plane \textbf{IP}.  \cite{bib1}}
	\label{fig:1}
\end{figure}

The physical quantity directly measured in an interferometer is a phase
difference. The setup is therefore arranged so that the difference of total phases
between the two arms coincides with the difference of geometric phases. As shown
schematically in Fig.~\ref{fig:1}, a polarized neutron beam is split into two
sub-beams propagating along two arms of different lengths, $l_a$ and $l_b$, and
subjected to magnetic fields of different magnitude, $B_a$ and $B_b$, oriented
along the same direction. The two beams are finally recombined at the
interference plane \cite{bib1}.
For each arm $J=a,b$, the magnetic field fixes the quantities
$
\theta_J
=
\frac{1}{2}
\arctan\!\left(
\frac{2\epsilon_{nn'}}{\delta m-|\mu_n B_J|}
\right),$ and
$
\Delta\omega_J=\omega_2(B_J)-\omega_1(B_J).
$
The dynamical phase difference can be canceled by choosing the arm lengths such
that
$
l_b
=
\frac{\Delta\omega_a\cos(2\theta_a)}
{\Delta\omega_b\cos(2\theta_b)}
\,l_a.
$
Under this condition, the residual phase observed at the interference plane is
purely geometric. Setting $t_a=l_a/v$, where $v$ is the neutron velocity, one
finds
\begin{align}
\Delta\Phi
=
\Delta\Phi^{g}
&=
\arg\!\left[
\cos\!\left(
\frac{\Delta\omega_a l_a}{2v}
\frac{\cos(2\theta_a)}{\cos(2\theta_b)}
\right)
+
i\sin\!\left(
\frac{\Delta\omega_a l_a}{2v}
\frac{\cos(2\theta_a)}{\cos(2\theta_b)}
\right)
\cos(2\theta_b)
\right]
\nonumber\\
&\quad
-
\arg\!\left[
\cos\!\left(
\frac{\Delta\omega_a l_a}{2v}
\right)
+
i\sin\!\left(
\frac{\Delta\omega_a l_a}{2v}
\right)
\cos(2\theta_a)
\right].
\end{align}
It
vanishes in the absence of $n$--$n'$ mixing and depends on the two mirror-sector
parameters $\epsilon_{nn'}$ and $\delta m$, as well as on the externally controlled
magnetic fields and arm lengths \cite{bib1}.

\begin{figure}[htpb]
	\centering
	\includegraphics[width=0.5\linewidth]{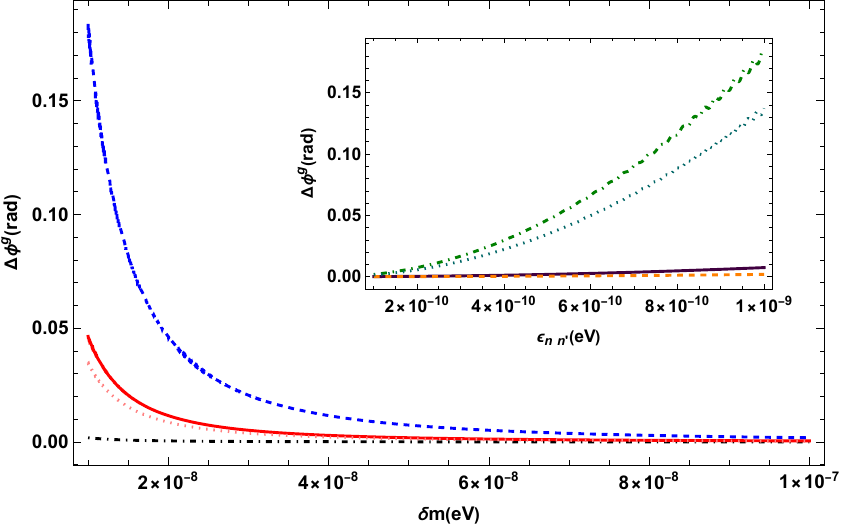}
	\caption{Geometric phase difference $\Delta\phi^{g}$ as a function of the mass
	splitting $\delta m$ for $\epsilon_{nn'} = 10^{-10}\,\mathrm{eV}$ (black dotted
	line), $\epsilon_{nn'} = 5\times10^{-9}\,\mathrm{eV}$ (red solid line), and
	$\epsilon_{nn'} = 10^{-9}\,\mathrm{eV}$ (blue dashed line). The inset shows
	$\Delta\phi^{g}$ as a function of $\epsilon_{nn'}$ for
	$\delta m = 10^{-8}\,\mathrm{eV}$ (green dotted line),
	$\delta m = 5\times10^{-8}\,\mathrm{eV}$ (purple solid line), and
	$\delta m = 10^{-7}\,\mathrm{eV}$ (orange dashed line). The parameters are
	$\lambda = 10\,\mathrm{\AA}$, $B_a = 0\,\mathrm{T}$,
	$B_b = 2\times10^{-4}\,\mathrm{T}$, and $l_a = 20\,\mathrm{cm}$. The dotted
	pink and cyan lines include the effect of the Earth's magnetic field,
	$B_a = 5\times10^{-5}\,\mathrm{T}$, for
	$\epsilon_{nn'} = 5\times10^{-10}\,\mathrm{eV}$ (pink) and
	$\delta m = 10^{-8}\,\mathrm{eV}$ (cyan).}
	\label{fig:3}
\end{figure}

The behavior of the geometric phase difference is shown in
Fig.~\ref{fig:3}. For fixed magnetic fields and arm length, the phase becomes
larger when the mass splitting $\delta m$ decreases and when the mixing amplitude
$\epsilon_{nn'}$ increases. This behavior follows directly from the mixing angle,
which is enhanced for larger $\epsilon_{nn'}$ and for values of
$|\mu_n B_J|$ approaching $\delta m$. The same figure also shows that the
Earth's magnetic field can affect the phase and must therefore be included or
controlled in a realistic implementation.

The proposed interferometric scheme is therefore sensitive to the existence of
neutron--mirror-neutron mixing through a purely geometric phase difference. The measurement of $\Delta\Phi^{g}$ does not
uniquely determine both $\epsilon_{nn'}$ and $\delta m$, but it can reveal the
presence of mixing and constrain the allowed parameter space when combined with
independent information \cite{bib1}.

\section{Conclusions}
\label{sec:conclusions}

We have shown that fermion--fermion interactions mediated by axion-like particles
can induce a geometric phase in neutron beams, which becomes observable once the
ordinary magnetic dipole--dipole contribution is canceled modulo $2\pi$. We have
also considered neutron--mirror-neutron mixing and shown that the coupling to
hidden mirror states generates a geometric phase in the ordinary neutron state.
Since this phase vanishes in the absence of ordinary--mirror neutron mixing, it
provides a distinctive interferometric signature of the hidden sector.

These examples show that geometric phases constitute robust and highly sensitive
observables for probing weak hidden-sector effects in quantum interferometric
systems. In particular, neutron interferometry may offer a complementary route to
standard searches for axion-like particles and mirror-matter candidates. A realistic
implementation will require precise control of magnetic-field inhomogeneities,
beam coherence, polarization effects, and environmental magnetic backgrounds.
A detailed analysis of these experimental aspects, including sensitivity estimates
and possible sources of systematic uncertainty, will be presented in a forthcoming
work.

\section*{Acknowledgements}
We acknowledge partial financial support from MUR and INFN, A.C. also acknowledges the COST Action CA1511 Cosmology and Astrophysics Network for Theoretical Advances and Training
Actions (CANTATA).

\section*{ORCID}

\noindent Antonio Capolupo - \url{https://orcid.org/0000-0002-8745-2522}

\noindent Gabriele Pisacane - \url{https://orcid.org/0009-0006-6626-6655}

\noindent Raoul Serao - \url{https://orcid.org/0000-0001-9507-240X}

\appendix

\end{document}